%% file: DPF2013-proceeding-wireless-Himansu.tex
\documentclass[12pt]{article}
\usepackage{graphicx}
\usepackage{color}

\newcommand\pubnumber{DPF2013-221}
\newcommand\pubdate{\today}

\def\napoli{High Energy Physics Division\\
Argonne National Laboratory, Argonne, IL-60439, USA}
\def\niu{Electrical Engineering Department\\
Northern Illinois University, Dekalb, IL-60115, USA}
\def\support{\footnote{
Presented by Himansu Sahoo at the DPF 2013 Meeting of the American Physical Society Division of Particles and Fields, Santa Cruz, California, August 13-17, 2013.}}

\def\Title#1{\begin{center} {\Large #1 } \end{center}}
\def\Author#1{\begin{center}{ \sc #1} \end{center}}
\def\Address#1{\begin{center}{ \it #1} \end{center}}

\newcommand\pubblock{\rightline{\begin{tabular}{l} \pubnumber\\
         \pubdate  \end{tabular}}}
\newenvironment{Abstract}{\begin{quotation}  }{\end{quotation}}
\newenvironment{Presented}{\begin{quotation} \begin{center} 
             PRESENTED AT\end{center}\bigskip 
      \begin{center}\begin{large}}{\end{large}\end{center} \end{quotation}}
\def\Acknowledgments{\bigskip  \bigskip \begin{center} \begin{large}
             \bf ACKNOWLEDGMENTS \end{large}\end{center}}

\input econfmacros.tex

\begin{document}
\begin{titlepage}
\pubblock

\vfill
\Title{Design and Testing of a Wireless Demonstrator for Large Instrumentation Systems}
\vfill
\Author{H. Sahoo\support, P. De Lurgio, Z. Djurcic, G. Drake, A. Kreps and M. Oberling}
\Address{\napoli}
\Author{R. Hashemian and T. Pearson}
\Address{\niu}
\vfill
\begin{Abstract}
In this proceeding, we report the development of a wireless demonstrator intended to readout instrumentation systems 
having thousands of channels.
A data acquisition system was designed and tested based on compliant 
implementation of 802.11n based hardware and protocols. 
This project is for large detectors containing photomultiplier tubes. 
Both free-space optical and radio frequency techniques were tested for wireless power transfer.
The front-end circuitry, including a high-voltage power supply was powered wirelessly, 
thus creating an all-wireless detector readout. 
The system was successfully tested as a single detector module, which was powered wirelessly and transmitted data wirelessly.
The performance of the prototype system and how a large scale implementation of 
the system might be realized are described in this proceeding.
\end{Abstract}
\vfill
\begin{Presented}
DPF 2013\\
The Meeting of the American Physical Society\\
Division of Particles and Fields\\
Santa Cruz, California, August 13--17, 2013\\
\end{Presented}
\vfill
\end{titlepage}
\def\thefootnote{\fnsymbol{footnote}}
\setcounter{footnote}{0}
%


\section{Motivation for Wireless DAQ}

In several areas of scientific research, the size and complexity of detectors have become exceedingly large. 
For example, detectors used in Nuclear and Elementary Particle Physics can have dimensions of the order of 10-100 meters and contain thousands to millions of readout channels. 
This create a significant challenge to power the electronics as well as to transfer data.
Traditional approach of using electrical cables have become complicated and expensive at a larger scale.
Specially, cabling is not practical for detectors in remote location or in hostile environment.
To overcome these problems, we developed an alternative approach that uses wireless techniques to power and transmit data.
A stand-alone photomultiplier tube (PMT) base detector was designed and tested in free space, that operates from 
wireless power and then transfers data wirelessly. 
We carried out this case study for large detectors containing PMTs, which are commonly used in high energy physics research. 
The primary purpose is to ascertain the feasibility and 
practicality of such devices as single detector modules that can be configured as arrays in a large detector.
This approach has potential to eliminate the need of expensive and massive cable plants, to simplify the process of installation 
and repair, and to reduce the detector dead mass.


\section{Design Considerations}

We have explored different technologies, such as free-space optical and radio frequency (RF) to wirelessly transmit data and power.
But, we selected the technology that is inexpensive and off-the-self that can be easily implemented while meeting the performance
goals of our R\&D project.

For wireless data transmission, optical links support higher data rates (1 Gbit/s~\cite{optical-link}) than those use RF. 
However, RF transmission does not require line-of-sight, 
and an individual receiver (i.e. access point) can communicate with many front-ends. 
Since both of these advantages provide significant simplification and cost reduction, 
RF data transmission was chosen for this project.
We focused on wireless local area network (WLAN) technologies with 802.11n variant,
because it offers the highest data throughput and has sufficient range for data transmission.
For instance, a single steam 802.11n link has a total data rate of approximately 65 Mbit/s~\cite{connect-blue}.
While this is sufficient for our single prototype front-end, it provides
greater challenge for large detectors in transferring data 
from thousands of readout channels over a limited and common frequency spectrum.
We addressed this problem by filling the available frequency space with many access points, 
as described in Ref.~\cite{wireless-paper}.
One of the frequency range in 802.11n is centered around 5.5 GHz with an overall bandwidth 
of approximately 1.2 GHz (4.9$-$6.1 GHz).
Single stream 802.11n access points can have an individual operating bandwidth of 20 MHz. 
This enabled us to populate up to 48 access points with the usable 1.2 GHz frequency spectrum,
each communicating to nearly 64 PMT wireless read-outs or front-ends (a total of 3072 front-ends).
This design provided an overall data throughput of 1.68 Gbit/s. 

\begin{figure}[htbp]
\begin{center}
        \resizebox{7cm}{5cm}{\includegraphics{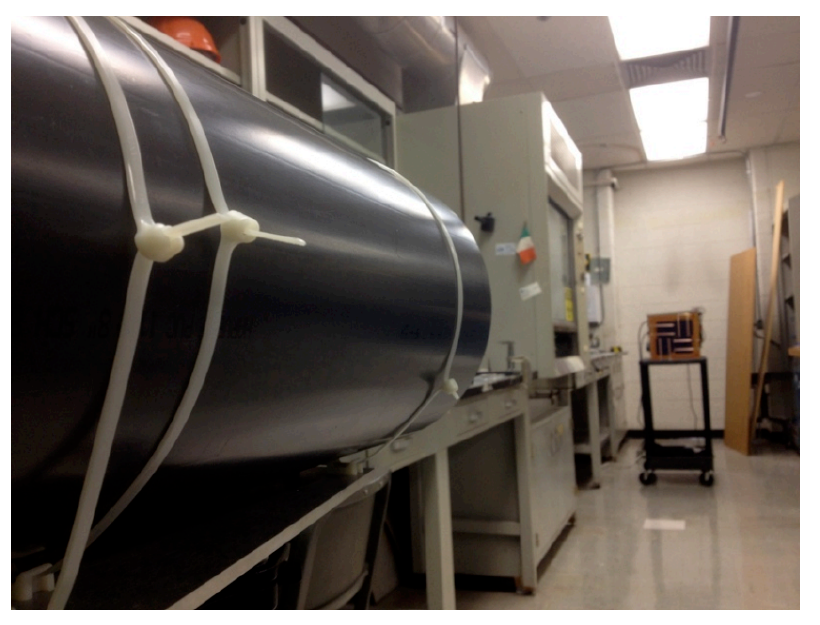}}
                \resizebox{0.5\columnwidth}{!}{\includegraphics{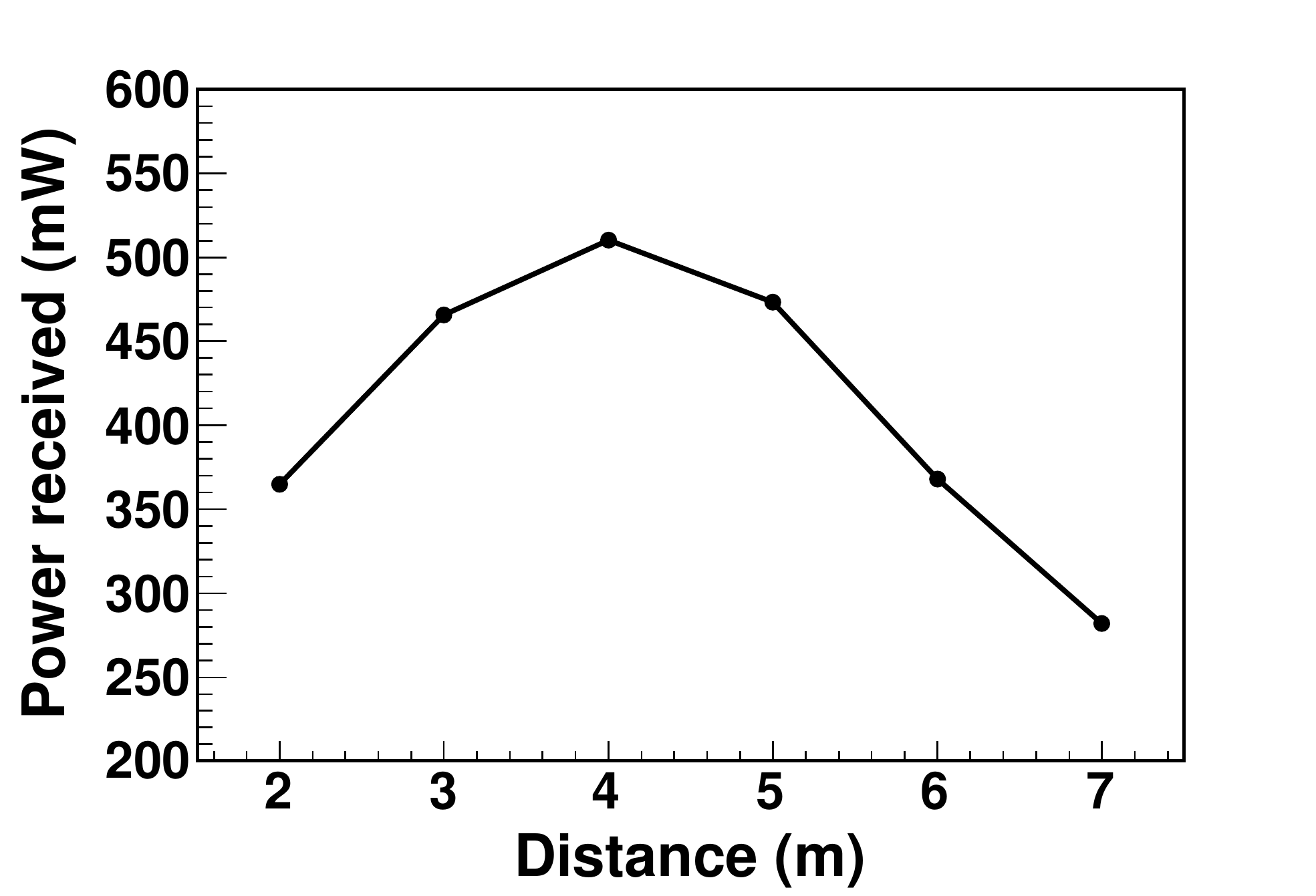}}
        \caption{
        	Apparatus for optical power transmission. Left figure shows the tube containing LED and lens, 
	and the photovoltaic receiver at the far end.
         	 Right figure shows the power received (mW) by the photovoltaic panel 
           as a function of distance (m) from the optical source. 
                     }
\label{fig:figure1-optical} 
\end{center}
\end{figure}

For wireless power transmission, we tested both optical and RF power transfer methods. 
The optical power demonstrator utilizes a high power light-emitting diode (LED)
that is collimated into an 8 inch diameter beam and is received by a photovoltaic (PV) panel, 
as shown in Fig.~\ref{fig:figure1-optical}. 
The LED is an OSRAM SFH 4751 with $3.5$ W optical output, operated at a maximum DC current of 1 A. 
The LED wavelength is $940$ nm which 
matches the peak efficiency of the Delsolar $156 \times 156 \; \mathrm{mm}^2$ photovoltaic cell used in our 
$312 \times 280 \; \mathrm{mm}^2$ PV panel array. 
This test system met our requirements of receiving 250 mW of power at 5 meters. 

\begin{figure}[ht]
\begin{center}
        \resizebox{7cm}{5cm}{\includegraphics{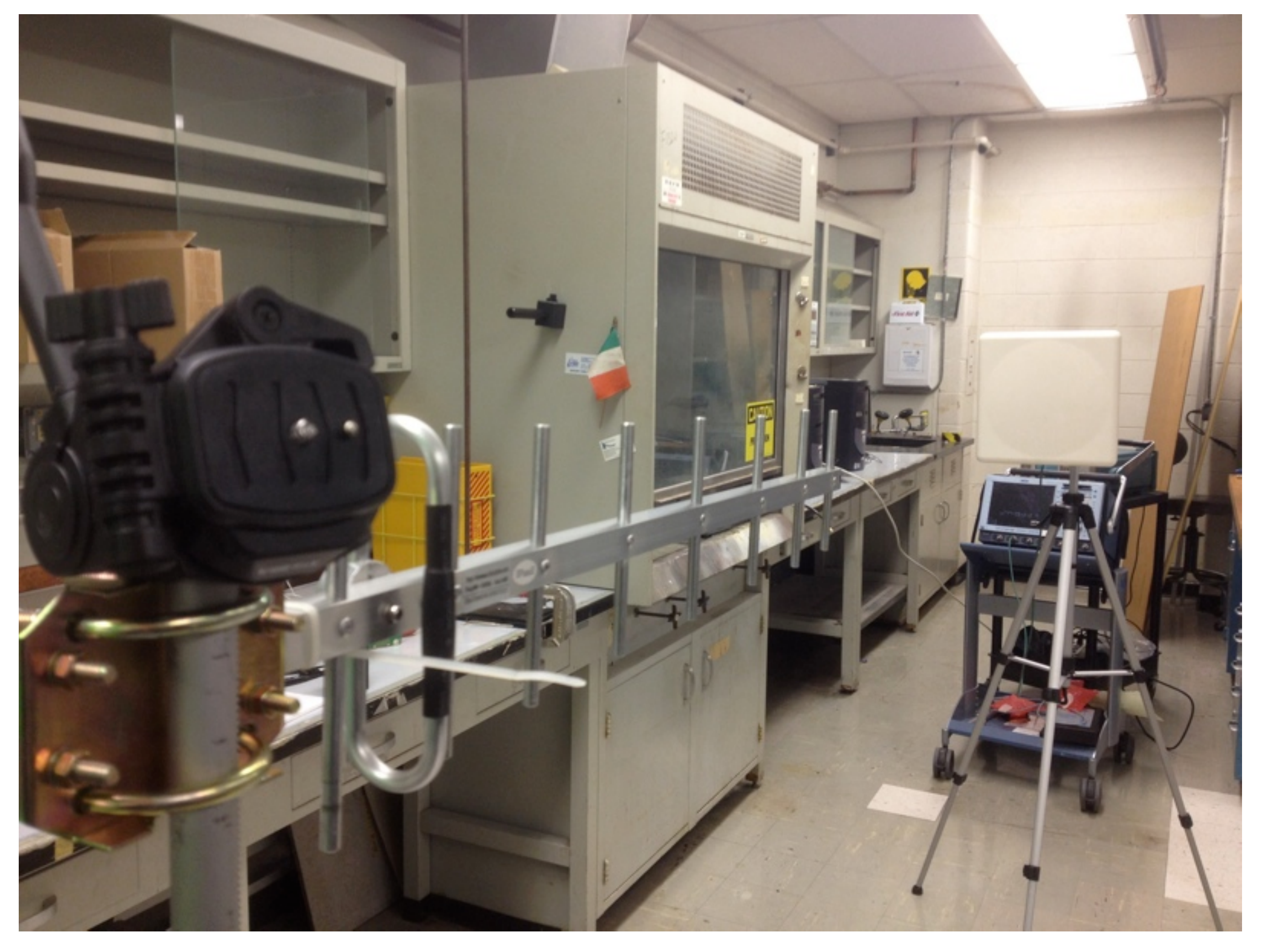}}
                \resizebox{0.5\columnwidth}{!}{\includegraphics{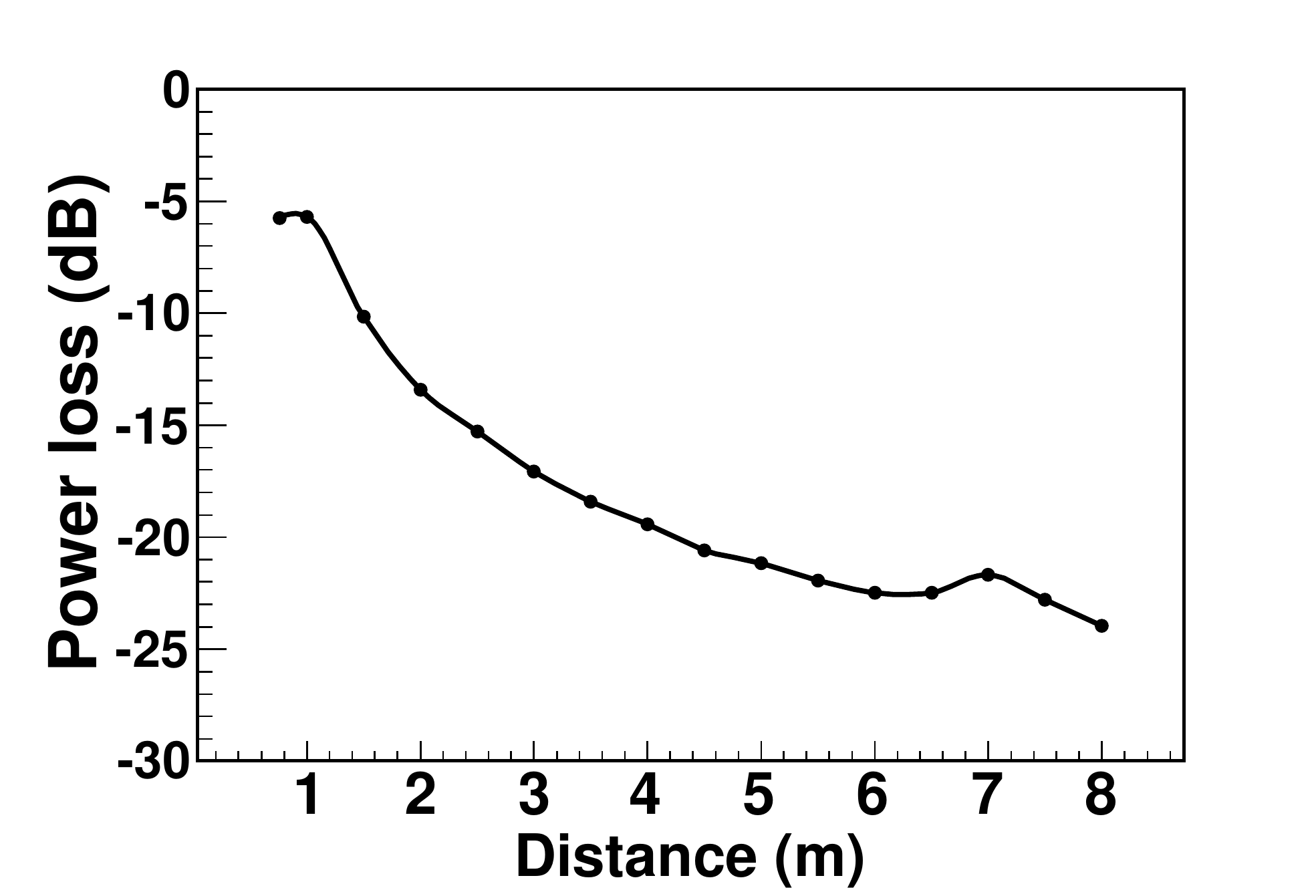}}
        \caption{
        Apparatus for RF power transmission.
        Left figure shows the RF transmitter ($14$ dBi gain Yagi antenna) in the foreground, 
        with the RF receiver ($11$ dBi gain patch antenna) at the far end.
        Right figure shows the power loss (dB) as a function of distance (m) from the source. 
        }
\label{fig:figure2-RF} 
\end{center}
\end{figure}

The RF power demonstrator uses high-gain directional microwave antennas.
The setup consisted of a function generator driving a 14 dBi gain Yagi antenna at 915 MHz 
with an output power of 10 dBm, which was received by a 11 dBi gain patch antenna, as shown in Fig.~\ref{fig:figure2-RF}. 
The power received was measured in free space to minimize scattering from surrounding objects.
As the transmission distance increased, the power received fell rapidly; the power loss being 20 dBm at 5 meters.
Thus, a 25 W source will be required to receive our targeted 250 mW power.

For this feasibility study, we chose the optical method to implement in our prototype. 
It provided a DC source at the receiver end, which is relatively easy to utilize. 
In order to transmit RF power, it has to be converted into a DC supply at the receiver end,
which is commercially available for only a 100 mW power~\cite{rf-dc-convert}. 
However, RF power could be a better choice for a large detector 
because one source can power many front-ends, thereby reducing the complexity and cost of the system.


\section{Wireless Data Acquisition System}

\begin{figure}[ht]
\begin{center}
                          \resizebox{3.5cm}{!}{\includegraphics{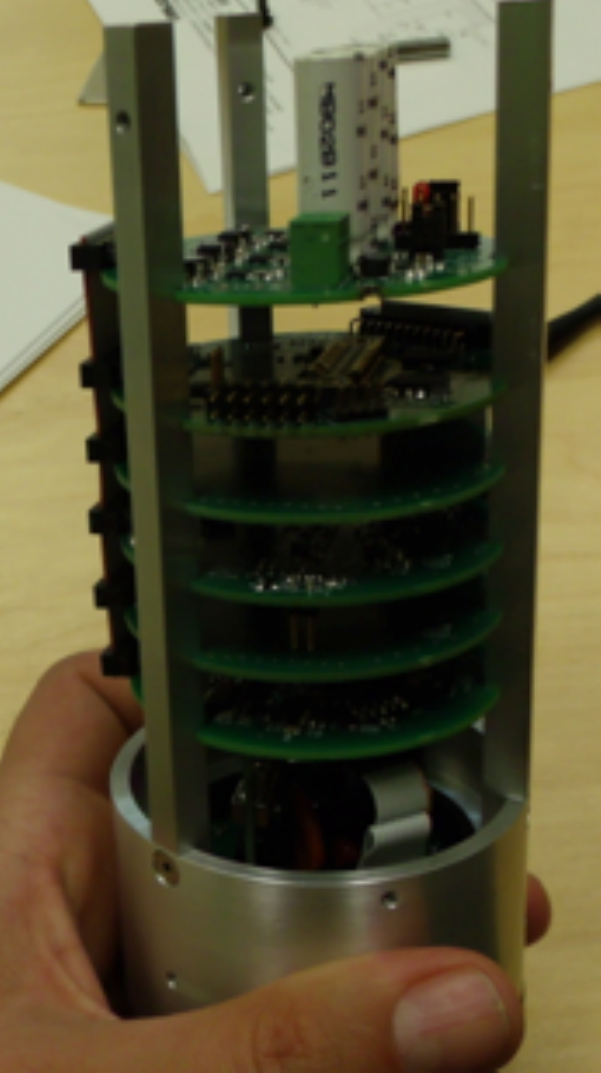}}
\hspace{2cm}  \resizebox{8cm}{!}{\includegraphics{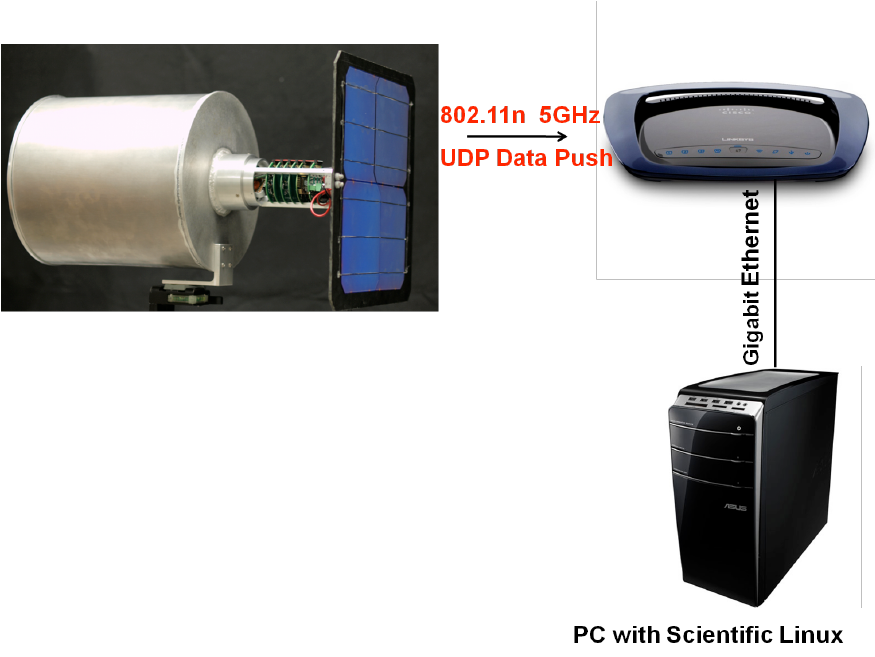}}
        \caption{
 		The printed circuit boards used in the wireless prototype system (left) and configuration of the 
		prototype readout system (right). 
        }
\label{fig:figure3-prototype}
\end{center}
\end{figure}

The wireless prototype system is comprised of four boards: 
a power board, which receives wireless power and generates different voltages needed by the system;
a digital board, which processes the data and does wireless data transmission;
a front-end board, which does shaping and digitization of the PMT signal; 
and a high voltage board, which generates high voltage for the PMT.
The high voltage board uses a standard Cockroft-Walton (CW) switching circuit to boost the 24 V input voltage 
up to 2000 V, which is needed for the 10-stage tube PMT.
We chose a large 10 inch diameter PMT, a Hamamatsu R7081HQE~\cite{R7081HQE}, which has dark noise rate of $\sim$10 kHz. 
All the boards are physically arranged inside a tube as shown in Fig.~\ref{fig:figure3-prototype}.
One end of the tube was fitted into the base of the PMT and the other end was connected to a photovoltaic panel.
Details of the wireless demonstrator are described in Ref.~\cite{wireless-paper}.

\begin{figure}[ht]
        \resizebox{0.5\columnwidth}{!}{\includegraphics{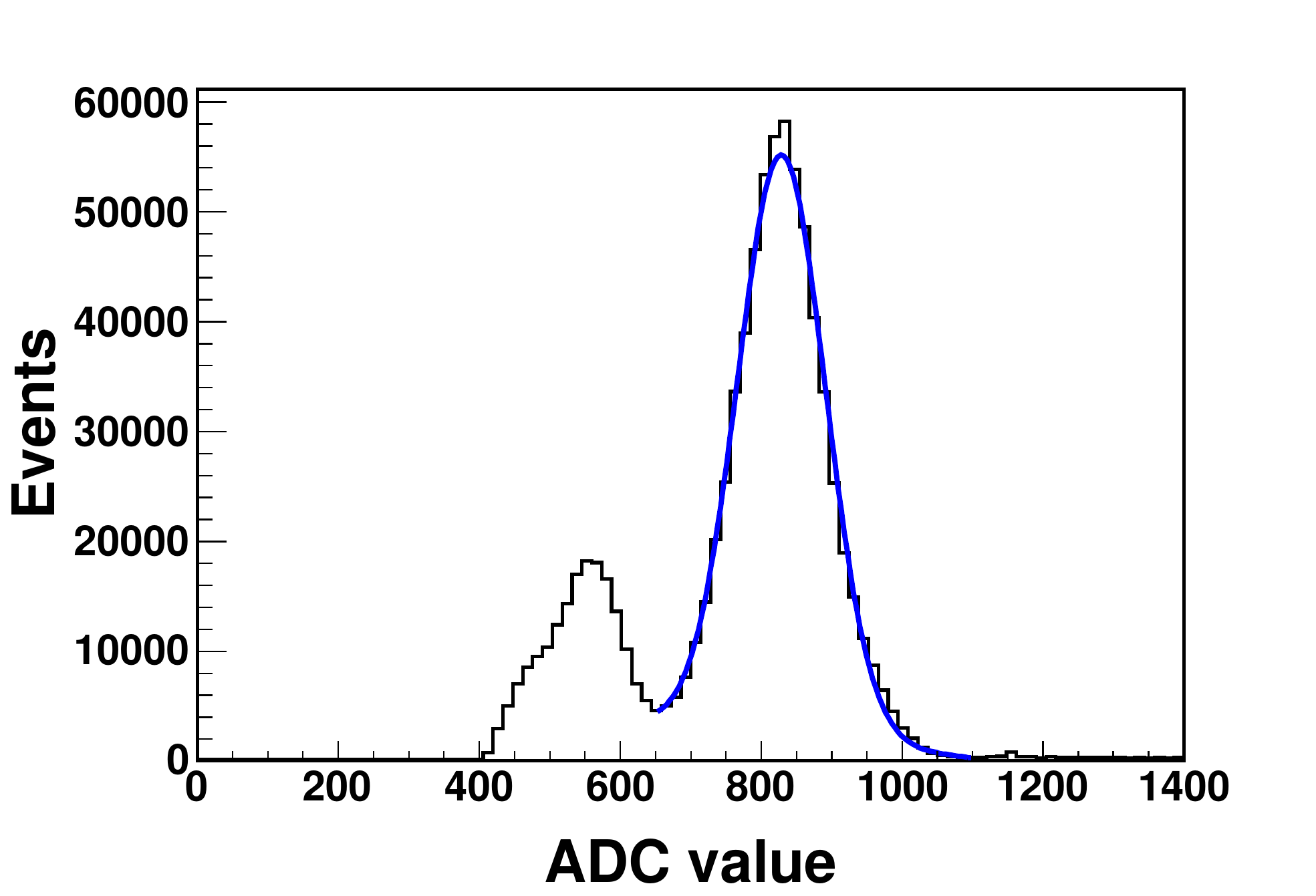}}
       \resizebox{0.5\columnwidth}{!}{\includegraphics{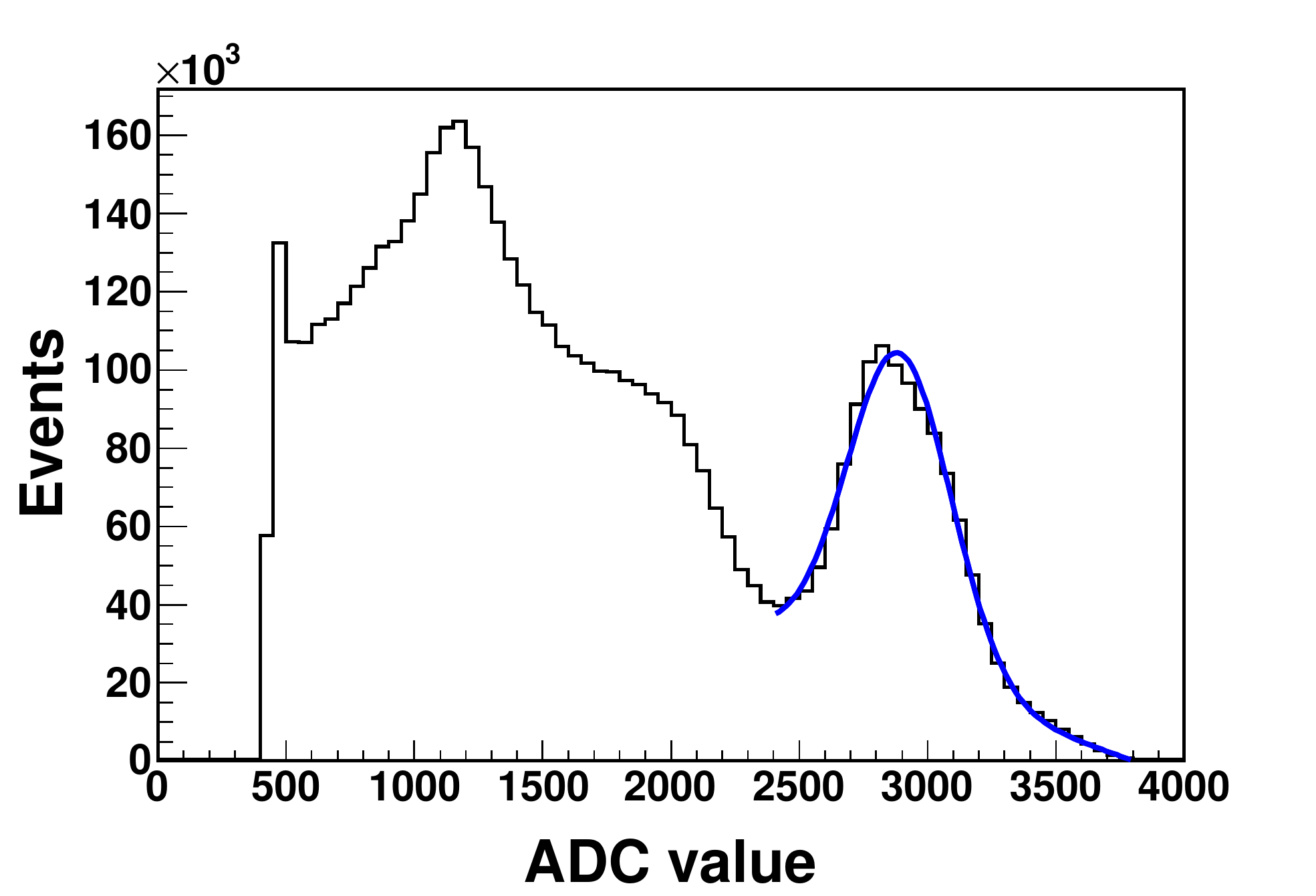}}
        \caption{Response from 60 keV $^{241}$Am source (left) and from 661.7 keV $^{137}$Cs source (right).  PMT HV in on and system is powered wirelessly and wireless data readout. }
\label{fig:figure4-tests} 
\end{figure}

The prototype readout system consisted of a Scientific Linux computer 
and a Cisco E3000~\cite{cisco-access} running DD-WRT firmware as an access point.
The front-end transmitted data wirelessly once per second as a single UDP packet using 802.11n in the 5 GHz band. 
A readout program running on the server received and stored the incoming UDP packets. For each asynchronous PMT trigger,
we stored the pulse height (2 bytes) and time stamp (4 bytes) information.
We have collected data with the prototype system operated from wireless power 
(using the optical source and received by the photovoltaic panel)
and with wireless data readout. 
The performance achieved along with target specifications are summarized in Table~\ref{tab:table1-summary}.
The system is capable of sending greater than 10k events/s.
To test the data acquisition capability, a sodium iodide crystal was attached to the PMT and tested 
with $^{241}$Am and $^{137}$Cs sources. 
The $^{137}$Cs data yields a $17\%$ energy resolution with sodium iodide.
The response from the sources are shown in Fig.~\ref{fig:figure4-tests}. 

\begin{table}[htbp]
\begin{center}
\caption{Summary of the initial goals and achieved performance for the wireless data acquisition system.}
\begin{tabular}{c|c|c}
     \hline\hline
Specification                                                  & Target                        & Performance   \cr   \hline
Total power consumption (@ $10$ kHz)     &   $250$ mW                   & $386$ mW           \cr  
\hspace{1cm} Digital                                    &   $120$ mW                   & $216$ mW          \cr
\hspace{1cm} Front-end                              &    $30$ mW                    &  $39$ mW          \cr
\hspace{1cm} HV                                          &    $80$ mW                    & $131$ mW          \cr   \hline
Maximum event rate                                    &    $10$ kHz                    &   $80$ kHz          \cr    \hline
Data transfer rate                                          &    $35$ Mb/s                   &   $11$ Mb/s          \cr    \hline
Bit Error Rate                                                 &  $<$ 1$\times$10$^{-12}$ & Dropped Packets  \cr    \hline\hline
\end{tabular}
\label{tab:table1-summary} 
\end{center}
\end{table}

\section{Summary}

We successfully built and tested a wireless demonstrator, which was implemented in a photomultiplier 
tube base and received power and transmitted data wirelessly. 
The system transmitted data at a rate of 11 Mbit/s, which can support up to 16 front-ends per wireless channel.
While the power consumption was slightly greater than our target, it was still low enough to allow the system to operate from our optical power system. 
In the longer-term, we intend to implement RF power transfer to facilitate the simplification by using 
one transmitter to power many receivers. 
Additionally, we will also investigate the use of a custom ASIC for lower power 
operation of the front-end and Cockroft-Walton control circuitry for a larger system.

\Acknowledgments
We acknowledge the support of Laboratory Directed Research \& Development funding from 
Argonne National Laboratory to carry out this project.
I thank the conference organizers for their invitation to present this work at DPF conference.

\end{document}

%% file: econfmacros.tex
\def\beq{\begin{equation}}
\def\eeq#1{\label{#1}\end{equation}}
\def\eeqn{\end{equation}}

\def\beqa{\begin{eqnarray}}
\def\eeqa#1{\label{#1}\end{eqnarray}}
\def\eeqan{\end{eqnarray}}

\let\bar=\overbar

\def\Dslash{\not{\hbox{\kern-4pt $D$}}}
\def\dslash{\not{\hbox{\kern-2pt $\del$}}}

\def\msb{{\bar{\ssstyle M \kern -1pt S}}}

